%% file: hammingball_arxiv.tex
\documentclass[11pt,a4paper]{article}

\usepackage[T1]{fontenc}
\usepackage[utf8]{inputenc}
\usepackage{authblk}
\usepackage{amsmath}
\usepackage{amsfonts}
\usepackage{amssymb}
\usepackage{graphicx}
\usepackage{bm}
\usepackage{bbm}
\usepackage{fullpage}
\usepackage[nomarkers,figuresonly]{endfloat}

\newcommand{\bfX}{{\bf X}}
\newcommand{\bfx}{{\bf x}}
\newcommand{\bfU}{{\bf U}}
\newcommand{\bfZ}{{\bf Z}}

\newcommand{\bfu}{{\bf u}}

\newcommand{\bfy}{{\bf y}}
\newcommand{\bfm}{{\bf m}}
\newcommand{\bfbeta}{{\bm \beta}}

\newcommand{\bfeta}{{\bm \eta}}
\newcommand{\bfw}{{\bf w}}
\newcommand{\bfz}{{\bf z}}
\newcommand{\bfzero}{{\bf 0}}

\newcommand{\dropcap}[1]{{#1}}

\author[1]{Michalis K. Titsias\thanks{mtitsias@aueb.gr}}
\author[1]{Christopher Yau\thanks{cyau@well.ox.ac.uk}}
\affil[1]{Department of Informatics, Athens University of Economics and Business, Athens, Greece}
\affil[2]{Wellcome Trust Centre for Human Genetics, University of Oxford, Roosevelt Drive, Oxford, United Kingdom}
\affil[3]{Department of Statistics, University of Oxford, 1 South Parks Road, Oxford, United Kingdom}

\begin{document}

\title{The Hamming Ball Sampler}

\maketitle

\begin{abstract}
We introduce the Hamming Ball Sampler, a novel Markov Chain Monte Carlo algorithm, for efficient inference in statistical models involving high-dimensional discrete state spaces. The sampling scheme uses an auxiliary variable construction that adaptively truncates the model space allowing iterative exploration of the full model space in polynomial time. The approach generalizes conventional Gibbs sampling schemes for discrete spaces and can be considered as a Big Data-enabled MCMC algorithm that provides an intuitive means for user-controlled balance between statistical efficiency and computational tractability. We illustrate the generic utility of our sampling algorithm through application to a range of statistical models. 
\end{abstract}

\input{introduction.tex}

\input{theory.tex}

\section{Results}

\input{results.tex}

\subsection{Tumor deconvolution through mixture modelling}

\input{tumor.tex}

\subsection{Sparse linear regression analysis}\label{sec:eqtl}

\input{varsel.tex}

\subsection{Factorial hidden Markov models}\label{sec:hmm}

\input{fhmm.tex}

\section{Discussion}

\input{discussion.tex}

\input{materials_arxiv.tex}

\section{Acknowledgments}
C.Y. is supported by a UK Medical Research Council New Investigator Research Grant (Ref. No. MR/L001411/1), the Wellcome Trust Core Award Grant Number 090532/Z/09/Z and the John Fell Oxford University Press (OUP) Research Fund. MKT acknowledges support from ``Research Funding at AUEB for Excellence and Extroversion, Action 1: 2012-2014".

\bibliographystyle{unsrt}
\bibliography{ref}

\end{document}

%% file: introduction.tex
\dropcap{S}tatistical inference of high-dimensional discrete-valued vectors or matrices underpins many problems across a variety of applications including language modelling, genetics and image analysis. Bayesian approaches for such models typically rely on the use of Markov Chain Monte Carlo (MCMC) algorithms to simulate from the posterior distribution over these objects. The effective use of such techniques requires the specification of a suitable proposal distribution that allows the MCMC algorithm to fully explore the discrete state space whilst maintaining sampling efficiency. While there have been intense efforts to design optimal proposal distributions for continuous state spaces, generic approaches for high-dimensional discrete state models have received relatively less attention but some examples include the classic Swendsen-Wang algorithm \cite{swendsen1987nonuniversal} for Ising/Potts models and more recent Sequential Monte Carlo methods \cite{schafer2013sequential}.

In this paper we consider Bayesian inference using MCMC for an unobserved latent discrete-valued discrete sequence or matrix $\bfX \in {\cal X}$, where each element $x_{ij} \in \{ 1, \dots, S \}$, given observations $\bfy = [ y_1, \dots, y_N ]$. We will assume that the observations are conditionally independent given $\bfX$ and model parameters $\theta$ so that the  joint distribution factorizes as   
$
	p(\bfy, \bfX,\theta) = \left [ \prod_{i=1}^N p(y_i|\bfX,\theta) \right ] p(\bfX,\theta).
$
We further assume that the posterior distribution $p(\bfX,\theta | \bfy)$ has a complex dependence structure so that standard MCMC schemes, such as a (Metropolis-within) Gibbs Sampler, using 
\begin{align}
	\theta & \leftarrow p(\theta|\bfX, \bfy), \label{eq:idealTheta} \\
	\bfX & \leftarrow p(\bfX|\theta,\bfy), \label{eq:idealX} 
\end{align}
or a marginal Metropolis-Hastings sampler over $\theta$ based on 
\begin{align}
	\theta & \leftarrow p(\theta|\bfy) \propto \sum_{\bfX \in {\cal X}} p(\bfy, \bfX,\theta),
	\label{eq:idealMarginalTheta}
\end{align}
are both intractable because exhaustive summation over the entire state space of $\bfX$ has exponential complexity. 

A popular and tractable alternative is to employ block-conditional (Metropolis-within) Gibbs sampling  in which subsets $\bfx_i$ of $\bfX$ are updated conditional on other elements being fixed using 
\begin{align}
	\theta & \leftarrow p(\theta|\bfX, \bfy), \label{eq:theta_blockGibbs} \\
	\bfx_i & \leftarrow p(\bfx_i|\bfX_{-i},\theta,\bfy), \forall i \label{eq:X_blockGibbs},  
\end{align}
where $\bfX_{-i}$ denotes the elements excluding those in $\bfx_i$. Typical block structures might be rows/columns of $\bfX$, when it is a matrix, or sub-blocks when $\bfX$ is a vector. Whilst block-conditional sampling approaches are often convenient (they may be of closed form allowing for Gibbs sampling without resort to Metropolis-Hastings steps), in high dimensions, major alterations to the configuration of $\bfX$ maybe difficult to achieve as this must be done via a succession of small (possibly low probability) incremental changes. Conditional sampling may lead to an inability to escape from local modes in the posterior distribution particularly if the elements of $\bfX$ exhibit strong correlations with each other and together with $\theta$. 

To address these problems, we propose a novel and generic MCMC sampling procedure for high-dimensional discrete-state models, named the ``Hamming Ball Sampler". This sampling algorithm employs auxiliary variables that allow  iterative sampling from slices of the model space. Marginalization within these model slices is computationally feasible and, by using sufficiently large slices, it is also possible to make significant changes to the configuration of $\bfX$. The proposed sampling algorithm spans a spectrum of procedures that contains the marginal and block-conditional Gibbs sampling strategies as extremes. At the same time, it allows the user to express many more novel schemes so that to select the one that best balances statistical efficiency and computational tractability. In this sense, the Hamming Ball Sampler is an example of a \emph{Big Data}-enabled MCMC algorithm. 

We demonstrate the utility of the sampling procedure with three different statistical models where exhaustive enumeration is impossible for realistic data sets and illustrate the considerable benefits over standard sampling approaches. 

\begin{figure*}	

	\centering
	\includegraphics[width=0.9\textwidth]{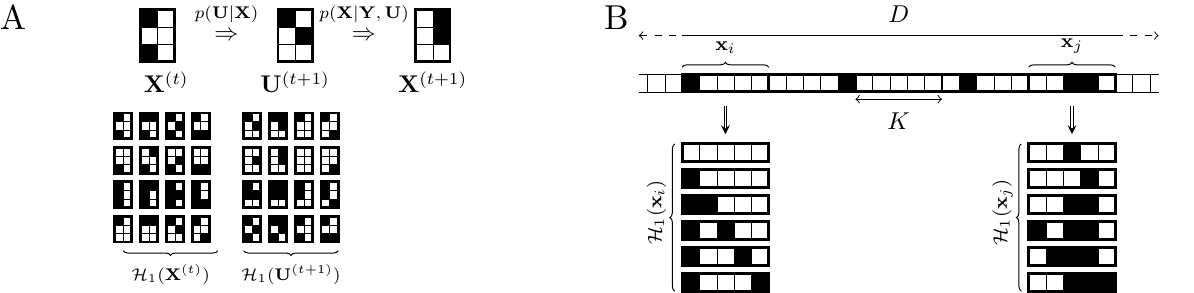}

\caption{Hamming Ball Sampler Illustration. Panel (A) illustrates a Hamming Ball update ($m=1$) for a $2 \times 3$ binary matrix $\bfX^{(t)}$ to $\bfX^{(t+1)}$ via $\bfU^{(t+1)}$ where the subsets $(\bfx, \bfu)$ correspond to columns of the matrix. Panel (B) illustrates a block strategy for the application of Hamming Ball sampling when $\bfX$ is a $D \times 1$ vector split into random blocks of size $K$.\label{fig:hammingball}}

\end{figure*}

%% file: theory.tex
\section{Theory}

\subsection{Construction} The Hamming Ball Sampler considers an augmented joint
 probability model that can be factorized as 
$p(\bfy, \bfX, \theta, \bfU) = p(\bfy, \bfX, \theta) p(\bfU|\bfX)$ where the extra factor $p(\bfU|\bfX)$ is a conditional distribution over an auxiliary variable $\bfU$ which lives in the same space and has the same dimensions as $\bfX$. The conditional distribution $p(\bfU|\bfX)$ is chosen to be an uniform distribution over a neighborhood set $\mathcal{H}_m(\bfX)$ centered at $\bfX$,  
$
	p(\bfU|\bfX) = \frac{1}{Z_m} \mathbb{I}(\bfU \in {\cal H}_m(\bfX)),
$ 
where $\mathbb{I}(\cdot)$ denotes the indicator function and the normalizing constant $Z_m$ is the cardinality of ${\cal H}_m(\bfX)$. 

The neighborhood set $\mathcal{H}_m(\bfX)$ will be referred to as a \emph{Hamming Ball} since it is defined through Hamming distances so that  
$
	{\cal H}_m(\bfX) = \{ \bfU :  d( \bfu_i, \bfx_i ) \leq m, i=1, \ldots,P \}.
$
Here, $d(\bfx_i, \bfu_i)$ denotes the Hamming distance $\sum_j \mathbb{I}(u_{ij} \neq x_{ij})$ and
the pairs $(\bfu_i,\bfx_i)$ denote non-overlapping subsets of corresponding entries in $(\bfU,\bfX)$ such that $\cup_{i=1}^P \bfu_i = \bfU$ and $\cup_{i=1}^P \bfx_i = \bfX$. Also, the parameter $m$ denotes the maximal distance or radius of each individual Hamming Ball set. For instance, these pairs can correspond to different matrix columns so that $\bfx_i$ will be the $i$-th column of $\bfX$ and $\bfu_i$ the corresponding column of $\bfU$. Hence, the Hamming Ball ${\cal H}_m(\bfX)$ would consist of all matrices whose columns are \emph{at most} $m$ elements different to $\bfX$.  

\subsection{Gibbs sampling} The 
principle behind the Hamming Ball Sampler is that the 
use of Gibbs sampling for the augmented joint probability distribution $p(\bfy, \bfX, \theta, \bfU)$ admits the \emph{target} posterior distribution $p(\bfX, \theta|\bfy)$ as a by-product (since marginalization over $\bfU$ recovers the target distribution). Specifically, the Hamming Ball Sampler alternates between the 
steps:  
\begin{align}
    \bfU & \leftarrow p(\bfU|\bfX), \label{eq:u_hbs} \\
	( \theta, \bfX ) & \leftarrow p(\theta, \bfX|\bfy, \bfU) \label{eq:X_hbs}.
\end{align}
The update of $(\theta, \bfX)$ can be implemented as two conditional (Gibbs) updates:
\begin{align}
	\theta & \leftarrow p(\theta| \bfX, \bfy),  \label{eq:theta_hbs} \\
	\bfX & \leftarrow p(\bfX|\theta, \bfU, \bfy) \label{eq:Xonly_hbs}.	
\end{align}
Or, alternatively, as a joint update via a Metropolis-Hastings accept-reject step that draws a new $(\theta',\bfX')$ 
from the proposal distribution  $Q(\theta',\bfX'|\theta,\bfX) = p(\bfX'|\theta', \bfU, \bfy) q(\theta'|\theta)$ and accepts it with probability
$
	\min \left ( 1, \frac{ p(\theta', \bfU,\bfy) q(\theta|\theta') }{ p(\theta,\bfU,\bfy) q(\theta'|\theta) } \right ),
$
where $q(\theta'|\theta)$ is a proposal distribution over the model parameters. 

The above algorithms consist of generalizations of the (Metropolis-within) Gibbs 
and marginal schemes outlined in \eqref{eq:idealTheta}-\eqref{eq:idealX} and \eqref{eq:idealMarginalTheta} 
since the latter algorithms are obtained as special cases when the radius $m$ becomes large enough 
(see {\it SI A: Further details about the Hamming Ball Sampler}). 

\subsection{Restricted state space} Crucially, the restricted state space defined by the Hamming Ball, that has been injected into the model 
via the auxiliary factor  $p(\bfU|\bfX)$, means that the conditional distribution $p(\bfX|\theta,\bfU,\bfy)$ can be tractably computed as 
\begin{equation}
p(\bfX|\theta,\bfU,\bfy) = \frac{p(\bfy, \bfX, \theta) \mathbb{I}(\bfX \in {\cal H}_m(\bfU))}
{p(\theta, \bfU,\bfy)}, 
\end{equation}
where $p(\theta, \bfU,\bfy) = \sum_{\bfX' \in \mathcal{H}_m(\bfU)} p(\bfy, \bfX', \theta)$ is the normalizing constant found by exhaustive summation over all admissible matrices inside the Hamming Ball $\mathcal{H}_m(\bfU)$. Through careful selection of $m$, the cardinality of ${\cal H}_m(\bfU)$ will be considerably less than the cardinality of  $\cal X$ so that exhaustive enumeration of all elements inside the Hamming Ball would be computationally feasible. 

Overall, the proposed construction uses the auxiliary variable $\bfU$ to define a {\em slice} of the model given by $\mathcal{H}_m(\bfU)$. Sampling of $(\theta, \bfX)$ is performed within this sliced part of the model through $p(\theta, \bfX|\bfy, \bfU)$. At each iteration, this model slice randomly moves via the re-sampling of $\bfU$ in step \eqref{eq:u_hbs}, which simply sets $\bfU$ to a random element from $\mathcal{H}_m(\bfX)$. This re-sampling step allows for {\em random exploration} that is necessary to ensure that the overall sampling scheme is ergodic. The amount of exploration depends on the radius $m$ so that $\bfU$ can differ from the current state of the chain, say $\bfX^{(t)}$, at most in $m P$ elements, i.e.\ the maximum Hamming distance between $\bfU$ and $\bfX^{(t)}$ is $m P$. Similarly, the subsequent step of drawing the new state, say $\bfX^{(t+1)}$, is such that at maximum $\bfX^{(t+1)}$ can differ from $\bfU$ in $m P$ elements and overall it can differ from the  previous $\bfX^{(t)}$ at 
most in $2 m P$ elements. From these observations we can conclude that a necessary condition for the algorithm to be ergodic is that $m > 0$. Figure \ref{fig:hammingball}A graphically illustrates the workings of the Hamming Ball Sampler.

\subsection{Selection of blocks}

The application of the Hamming Ball Sampler requires the selection of the subsets or blocks $\{\bfx_1, \ldots, \bfx_P\}$. This selection will depend on the conditional dependencies specified by the statistical model underlying the problem to be addressed. For some problems, such as the tumor deconvolution mixture model considered later, there may exist a natural choice for these subsets (e.g.\ columns of a matrix) that can lead to efficient implementations. For instance, under a suitable selection of blocks the posterior conditional $p(\bfX|\theta,\bfU,\bfy)$ could be fully factorized, i.e.\ $p(\bfX|\theta,\bfU,\bfy) = \prod_{i=1}^P p(\bfx_i | \theta,\bfu_i,\bfy$), or have a simple Markov dependence structure (as for the factorial hidden Markov model example considered later) so that exact simulation of $\bfX$ would be feasible. 
In contrast, for unstructured models, where $\bfX$ is just a large pool of fully dependent discrete variables (stored as a $D$-dimensional vector), we can divide the variables into randomly chosen blocks $\bfx_i,\ i=1,\ldots,P$, so that they have equal length $K = \text{length}(\bfx_i)$. In such cases, exact simulation from $p(\bfX|\theta,\bfU,\bfy)$ may not be feasible and instead we can use the Hamming Ball operation to sequentially sample each block. More precisely, this variant of the algorithm can be based on the iteration \eqref{eq:u_hbs},\eqref{eq:theta_hbs}-\eqref{eq:Xonly_hbs} with the only difference that the steps \eqref{eq:u_hbs} and \eqref{eq:Xonly_hbs} are now split into $P$ sequential conditional steps, 
\begin{align}
\bfu_i & \leftarrow p(\bfu_i|\bfx_i), \ \bfx_i \leftarrow p(\bfx_i | \bfX_{-i}, \theta, \bfu_i, \bfy), \forall i.	
\label{eq:blockHamming}
\end{align}
This scheme can be thought of as a {\it block Hamming Ball Sampler} which incorporates standard block Gibbs sampling (see iterations \eqref{eq:theta_blockGibbs}-\eqref{eq:X_blockGibbs}) as a special case obtained when the radius $m$ is equal to the block size $K$. In a purely block Hamming Ball scheme we will have $m<K$ and in general the parameters $(m,K)$ can be used to jointly control algorithmic performance (see \textit{SI: Fig.} 1). This scheme is illustrated in Figure \ref{fig:hammingball}B and used later in the regression application 
and discussed in further detail in {\it SI B.1: Blocking strategies}.

\subsection{Computational complexity} 

To find the time complexity of the Hamming Ball Sampler we assume for simplicity that 
either $p(\bfX|\theta,\bfU,\bfy)$ factorizes across the blocks or we use the block scheme 
in \eqref{eq:blockHamming}. Then, for $P$ blocks of size $K$, the computational 
complexity of the Hamming Ball Sampler scales with the Hamming radius $m$, block size $K$ and $P$
according to $O(M P)$ where
$
	M = \sum_{j=0}^m (S-1)^{j} { K \choose j }.
$
Contrast this with the block Gibbs Sampler which has computational complexity of $O(S^K P)$ 
(where $S^K = \sum_{j=0}^K (S-1)^{j} { K \choose j }$)
and it is only applicable for small values of block size $K$. On the other hand, 
Hamming Ball sampling is more flexible since it can allow to use much larger block sizes 
by controlling the computational cost through both $K$ and $m$. An ideal choice of 
the Hamming radius is $m=K/2$ since, as discussed previously, it is possible to change $2m$ elements per Hamming Ball sampling update so with $m=K/2$ it is possible to update all $K$ elements in each block in a single update. In the {\em Results} section we shall see this outcome empirically in actual applications, however, for large-scale problems, it may not be practical to use Hamming distances beyond $m=1,2,3$. Our simulations will show that even in these circumstances the Hamming Ball Sampler can still be advantageous by providing the flexibility 
to balance statistical and computational efficiency.

\subsection{Extensions}
 
A simple generalization of the algorithm is obtained by allowing block-wise varying Hamming maximal distances.   
If we assume a varying radius for the individual Hamming Balls, then the conditional distribution over $\bfU$ becomes uniform on the generalized Hamming Ball
$
{\cal H}_{\bfm}(\bfX) = \{ \bfU :  d( \bfu_i, \bfx_i ) \leq m_i, i=1, \ldots,P \},
$
where $\bfm = (m_1,\ldots,m_P)$ denotes the set of maximal distances for each subset of variables. Furthermore, we could allow $\bfm$, at each iteration, to be randomly drawn from a distribution $p(\bfm)$ (see {\it SI B.2: Randomness over the Hamming Ball radius} for further discussion).


Alternate auxiliary conditional distributions $p(\bfU|\bfX)$ are also permitted. For instance, a more general auxiliary distribution can have the form $p(\bfU|\bfX) \propto \exp( - \lambda d(\bfU,\bfX) ) \mathbb{I}(\bfU \in {\cal H}_m(\bfX))$, with $\lambda \geq 0$, which for $\lambda>0$ is non-uniform and places more probability mass towards the center $\bfX$ (see {\it SI B.3: Non-uniform auxiliary Hamming Ball distributions} for further discussion). 

%% file: results.tex
We now illustrate the utility of our Hamming Ball sampling scheme through its application to three statistical models that involve high-dimensional discrete state spaces. We motivate the selection of each model through real scientific problems. 

%% file: tumor.tex
Tumor samples are genetically heterogeneous and typically contain an unknown number of distinct cell sub-populations. Current DNA sequencing technologies ordinarily produce data that comes from an aggregation of these sub-populations thus, in order to gain insight into the latent genetic architecture, statistical modelling must be applied to deconvolve and identify the constituent cell populations and their mutation profiles.

In order to tackle this problem, \cite{zare2014inferring} and \cite{xu2015mad} adopt a statistical framework in which the set of unobserved mutation profiles can be described as a $K \times N$ binary matrix $\bfX$, where $x_{ki} = 0/1$ denotes absence/presence of a mutation in the $k$-th cell type for the $i$-th mutation, and parameters $\theta = \{ \theta_1, \dots, \theta_K \}$ denote the proportion of the tumor attributed to each of the $K$ cell types. The observed sequence data $\bfy$ depends on $(\theta, \bfX)$ through the mutant allele frequencies $\phi = \frac{1}{2} \theta^T \bfX$ and we would like to simulate from the posterior distribution $P(\theta, \bfX | \bfy)$ and explore the possible configurations of the mutation matrix $\bfX$ that are most compatible with the observed data.\footnote{In \cite{zare2014inferring} and \cite{xu2015mad} statistical inference was conducted using deterministic approaches and not Monte Carlo sampling. They use massive numbers of random initializations to overcome local modes in the posterior. As our interest is in full posterior characterization we do not compare with the point estimates given by these methods.}

We compared three posterior sampling approaches: (i) a conventional block Gibbs sampling strategy that proceeds by conditionally updating one column $\bfx_i$ at a time with the remaining columns $\bfX_{-i}$ and the weights $\theta$ fixed, (ii) a Hamming Ball-based sampling scheme where we define the Hamming Ball as the set of all matrices such that each column $\bfx_i$ is at most $m$ bits different from the corresponding column $\bfu_i$ of the auxiliary matrix $\bfU$ and (iii) a \emph{fully marginalized} sampling strategy where $\bfX$ was marginalized through exhaustive summation over all column configurations (note, this corresponds to the Hamming Ball Sampler with $m=K$). Our data examples was chosen to be sufficiently small so that the fully marginalized sampler was practical. We refer to \textit{Materials and Methods} for detailed derivations and implementations of the following data examples.

We considered a simulated data example, illustrated in Figure 2A, where the observed sequence data is generated so that it can be equally explained by two different latent genetic architectures. This is an interesting example as one configuration corresponds to a linear phylogenetic relationship between cell types and the other to a branched phylogeny and represent fundamentally different evolutionary pathways. For this example, we would expect an \emph{efficient} sampler to identify both configurations and to be able to move freely between the two during the simulation revealing the possibility of the existence of dual physical explanations for the observed data. 

\begin{figure}[b]
    \centering
    \includegraphics[width=0.5\textwidth]{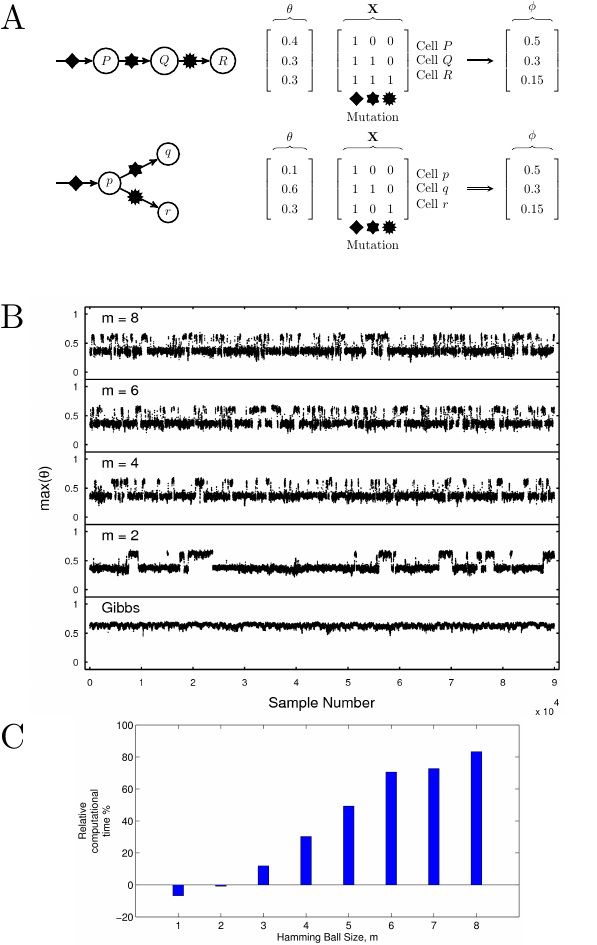}      
	\caption{Tumor deconvolution. (A) Two distinct clonal architectures that lead to the same mutant allele frequency vector $\phi = [ 0.5, 0.3, 0.15 ]'$. (B) Trace plots showing the sampled values of $\max(\theta)$. (C) Relative computational times for the Hamming Ball Sampler for various $m$ (times relative to the block Gibbs Sampler). }\label{fig:tumor-example}
\end{figure}

Figures 2B and C display trace plots of the largest component weight, $\max{(\theta)}$ and the relative computational times for the three sampling schemes (see also \textit{SI: Figs.} 2-3). All Hamming Ball samplers were effective at identifying both modes but the efficiency of the mode switching depends on the Hamming ball size $m$. This effectiveness can be attributed to the fact that the Hamming Ball schemes can \emph{jointly} propose to change up to $2mN$ bits across \emph{all} $N$ columns of the current $\bfX$. Furthermore, conditional updates of $\theta$ can be made by marginalizing over a range of mutation matrices. For $m \geq K/2$, the efficiency of the Hamming Ball Sampler is therefore close to the fully marginalized sampling procedure ($m=K=8$) but more computationally tractable if the number of mutations is large and exhaustive enumeration is impractical. The conditional updates employed by the block Gibbs Sampler requires significantly less computational effort but the approach is prone to being trapped in single posterior mode and our simulations show that it failed to identify the mode corresponding to the linear phylogeny structure ($\max{\theta} = 0.4$). In real application this could lead to incorrect scientific conclusions and we illustrate these potential impacts on a real cancer data set  in \emph{SI: C.2.  Tumor data analysis}.

%% file: varsel.tex
In this section we consider variable selection problems with sparse linear regression models. In this case, the high-dimensional discrete-valued object is a $D \times 1$ binary vector $\bfX$ whose entries are 1 when the corresponding covariate in the design matrix $\bfZ$ is associated with the response $\bfy$ (and zero otherwise) and $\theta$ consists of the regression and noise parameters (which can be analytically marginalized out in our chosen set-up, see {\em Materials and Methods}). The interest lies in the posterior distribution $P(\bfX|\bfZ, \bfy)$ which would inform us as to which covariates are most important for defining the observed response variable. Typically $\bfX$ is assumed to be sparse so that only a few covariates contribute to the explanation of the observations. These sparse linear regression models can arise in problems such as expression quantitative trait loci (eQTL) analysis which is concerned with the association of genotype measurements, typically single nucleotide polymorphisms (SNPs), with phenotype observations (see e.g. \cite{o2009review} for a review).

We began by simulating a regression dataset with $N=100$ responses and $D=1,200$ covariates in which there were two relevant covariates that fully explain the data while the reminder were noisy redundant inputs. These two covariates were chosen to be perfect \emph{confounders} so that only one of them is needed to explain the observed responses. As a consequence this sets up a challenging model exploration problem as only two out of $2^{1200}$ possible models represent the possible truth. We applied a range of MCMC sampling schemes to sample from the posterior distribution over this massive space of possible models including three block Gibbs Samplers (BG1-3), which conditionally update blocks of elements of $\bfX$ of size $K=1-3$ respectively, and three block Hamming Ball samplers (HB1-3) that use blocks of size $K=10$ and consider Hamming ball radii of $m=1-3$ respectively within each block. Note that the block Gibbs Samplers are special cases of the block Hamming Ball Sampler where $K=m$.

\begin{figure}[t]
	\centering
	\includegraphics[width=0.45\textwidth]{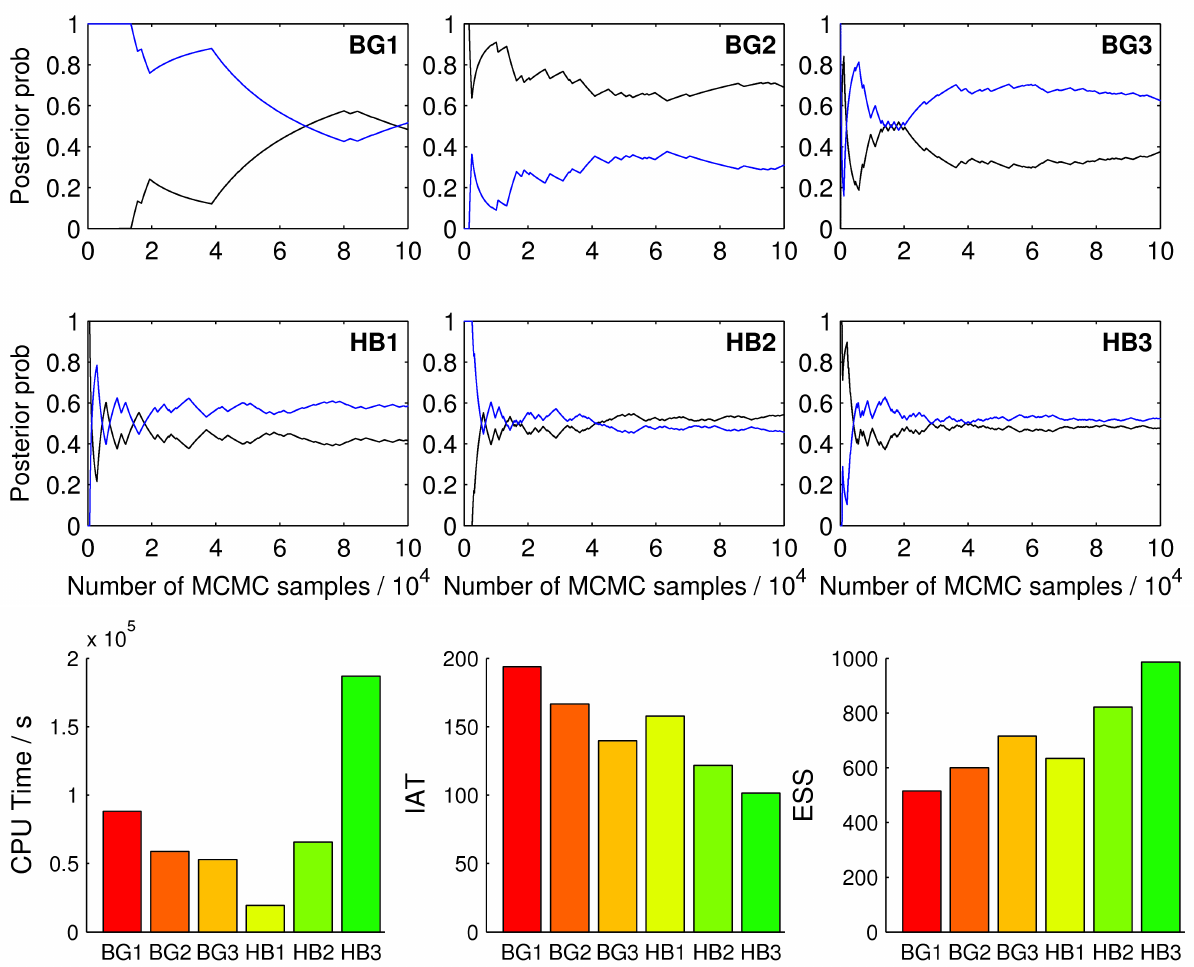}
	\caption{Comparison of block Gibbs and Hamming Ball sampling schemes for the simulation regression example. Top and middle rows give trace plots showing the running marginal posterior inclusion probabilities for $\bfx_{11}$ (black) and $\bfx_{611}$ (blue). Bottom row shows CPU times, integrated autocorrelation times (IAT) and effective sample size (ESS) estimates for each method.}
\label{fig:eQTL_toy}
\end{figure}

Figure 3 compares the relative performance of the various sampling schemes. The trace plots show the running marginal posterior inclusion probabilities of the two relevant variables $\bfx_{11}$ and $\bfx_{611}$ which converge to the expected values of 0.5 with the Hamming Ball Samplers but not with the block Gibbs Samplers. This indicates that the Hamming Ball schemes were mixing well, able to identify the two relevant variables and frequently switched between their inclusion. In contrast, the block Gibbs Samplers exhibited strong correlation effects (stickiness) that impaired their efficiency. 

For such a high-dimensional problem, the performance of the simplest Hamming Ball Sampler (HB1) was particularly outstanding as it used the least CPU time and achieved a lower integrated autocorrelation time than BG1 and BG2. The performance can be explained by the fact that the Hamming Ball sampling schemes can handle a large block of variables at a time but do not require exhaustive enumeration of all possible latent variable combinations within each block. This provides an important computational saving for sparse problems where most combinations will have low probability and the reason why the HB1 sampler was particularly effective for this example. In \emph{SI: D.3. Real eQTL analysis} we demonstrate the utility of this methodology for a real eQTL example involving 10,000 covariates.

%% file: fhmm.tex
We finally address the application of the Hamming Ball sampling scheme to the factorial hidden Markov model (FHMM) \cite{GhahramaniJordan97}. 
The FHMM is an extension of the hidden Markov model (HMM) \cite{rabiner1989tutorial} where multiple independent hidden chains run in parallel 
and cooperatively generate the observed data. The latent matrix $\bfX$ in this case represents a $K \times N$ $S$-valued discrete matrix, whose 
rows corresponds to $K$ hidden Markov chains of length $N$. Posterior inference in FHMMs is extremely challenging since it concerns the computation 
of $P(\bfX|\bfy,\theta)$ which comprises a fully dependent distribution in the space of the $S^{K N}$ possible configurations. This is an 
extraordinarily large space for even small values of $K$ and $N$.

Applications of FHMMs therefore frequently  
rely upon variational approximations or block Gibbs sampling which alternates between sampling 
a small set of rows of $\bfX$, conditional on the rest rows \cite{GhahramaniJordan97}. These Gibbs sampling schemes can easily become trapped in local modes due 
the conditioning structure which means major structural changes to $\bfX$ are unlikely to be proposed. Joint posterior updates 
can be achieved by applying the forward-filtering-backward-sampling algorithm (FF-BS) \cite{Scott2002}, with exhaustive enumeration, to simulate a 
sample from the posterior in $O(S^{2 K} N)$ time. 
However, although the use of FF-BS is quite feasible for even very large HMMs, it is only practical for very small values of $K$ and $N$ in FHMMs. 

We consider an additive FHMM with binary hidden chains which models the observation at time $i$ according to 
$\bfy_i = \sum_{k=1}^K x_{ki} \bfw_k  + \boldsymbol{\eta}_i$  where $\bfw_k$ is a parameter vector that describes the contribution of
the $k$-th feature when generating $\bfy_i$ (given that $x_{ki}=1$) and $\boldsymbol{\eta}_i$ is Gaussian noise, i.e.\  
$\boldsymbol{\eta}_i \sim \mathcal{N}({\bfzero, \sigma^2 I})$; see {\it Material and Methods}.
Such a model is useful for applications such as energy disaggregation, where an observed total electricity power at time instant $i$ is the 
sum of individual powers for all devices that are “on” at that time. We set up a simulated sequence of length $N=1000$ and $K=10$ 
hidden chains (simulation details can be found in {\it SI: E Factorial hidden Markov models}).  For all sampling schemes, we treat 
the feature contributions $\bfw_k$ as known parameters and we conduct inference on $\bfX$, that models presence/absence of these features,  
together with the noise variance parameter $\sigma^2$. We applied three block Gibbs Samplers (BG1-3) and three Hamming Ball-based sampling schemes (HB1-3). As in the tumor deconvolution example, 
the Hamming Ball is defined as the set of all matrices $\bfX$ such that each column $\bfx_i$ is at most $m$ bits different from the corresponding 
column $\bfu_i$ of the auxiliary matrix $\bfU$. We then applied FF-BS within the Hamming Ball to sample from the constrained posterior distribution 
$p(\bfX|\bfU, \bfy)$.

Figure 4 shows the utility of the different sampling schemes applied to the FHMM. Clearly, the Hamming Ball Samplers, and particularly the schemes HB2-3, are able to escape from local modes of the posterior distribution and sample values for $\bfX$ that have much higher posterior probability 
than values sampled by block Gibbs Samplers. The configurations of $\bfX$ sampled by the HB2-3 schemes are also 
close to the true $\bfX_{true}$ that generated the data. Further, the latter algorithms 
were able to correctly infer the level of the noise variance that generated the data while the block Gibbs Samplers and HB1 have inferred 
larger noise variances so that they wrongly explain some true signal variation as noise. Finally, we can conclude that HB2 is the scheme 
that best balances computational time and sampling efficiency 
since, while it has similar CPU time with the most advanced block Gibbs Samplers (BG2-3), it allocates computational resources very differently from the BG algorithms
which results in significant improvement of the sampling efficiency. 
In {\it SI: E.3. Energy disaggregation} we demonstrate the utility of Hamming Ball sampling 
to a real energy disaggregation data sequence of length $67,200$.
 
\begin{figure}
\centering
\begin{tabular}{c}
{\includegraphics[width=0.48\textwidth]
{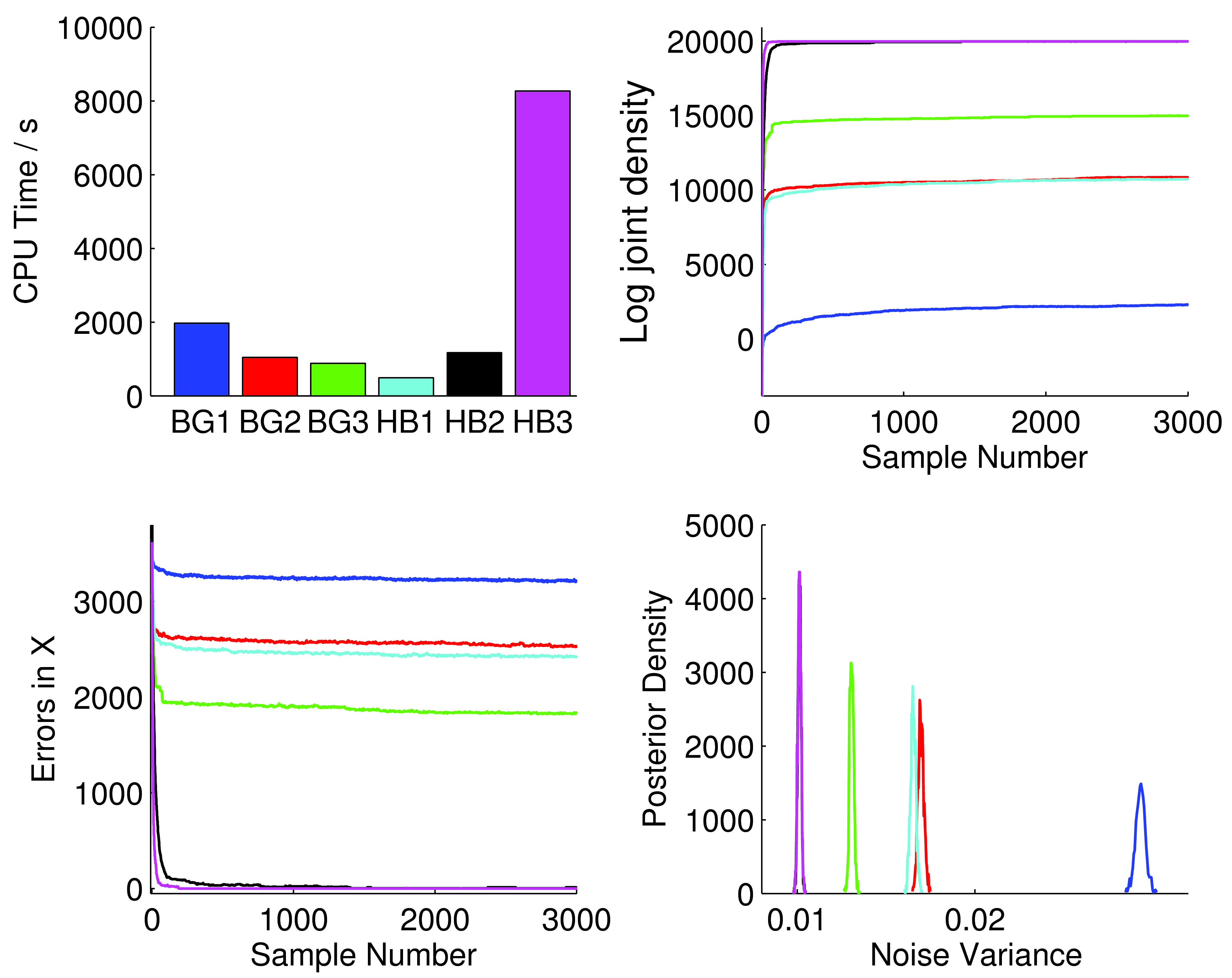}}
\end{tabular} 
\caption{Comparison of block Gibbs and Hamming Ball sampling schemes for the
simulation FHMM example. Top row shows CPU times and log joint density values during sampling. 
The bottom row shows the number of elements in $\bfX$ that differ from the true matrix $\bfX_{true}$ that generated the data and the Monte Carlo posterior densities over the inferred noise variance $\sigma^2$ (the true value was $0.01$).} 
\label{fig:fhmm_toy}
\end{figure}

%% file: discussion.tex
The Hamming Ball Sampler provides a generic sampling scheme for statistical models involving high-dimensional discrete latent state-spaces that generalizes and extends conventional block Gibbs sampling approaches. In our investigations, we have applied the Hamming Ball sampling scheme to three different statistical models and shown benefits over standard Gibbs samplers. Importantly, the Hamming Ball Sampler gives the statistical investigator control over the balance between statistical efficiency and computational tractability through an intuitive mechanism - the specification of the Hamming Ball radius and the block design strategy - which is important for Big Data applications. Yet, we have also demonstrated that in many problems, basic Hamming Ball samplers ($m=1$) that are computationally inexpensive can still give relatively good performance compared to standard block Gibbs sampling alternatives. 

Throughout we have provided pure and unrefined Hamming Ball sampler implementations. In actual applications, the proposed methodology should not be seen as a single universal method for speeding up MCMC but a novel addition to the toolbox that is currently available to us. For example, the computations performed within each Hamming Ball update are often trivially parallelizable which would allow the user to take advantage of any special hardware for parallel computations, such as graphics processing units \cite{suchard2010understanding,lee2010utility}. In addition, Hamming Ball sampling updates can also be used alongside standard Gibbs sampling updates as well as within parallel tempering schemes in Evolutionary Monte Carlo algorithms
 \cite{brooks2011handbook}. 

Finally, we believe the ideas presented here can have applications in many areas not yet explored, such as Bayesian nonparametrics (e.g.\ in the Indian Buffet Process) and Markov Random Fields. Further investigations are being conducted to develop the methodology for these statistical models.

%% file: materials_arxiv.tex
\section{Materials and Methods}
\subsection{Tumor deconvolution with mixture modelling} We give a description of the statistical model underlying the tumor deconvolution example in the following. We assume that the data $\bfy = \{ r_i, d_i\}_{i=1}^N$ consists of $N$ pairs of read counts where $r_i$ corresponds to the number of sequence reads corresponding to the variant allele at the $i$-th locus and $d_i$ is the total number of sequence reads covering the mutation site. The distribution of the variant allele reads given the total read count follows a Binomial distribution
$
	r_i \sim \mathrm{Binomial}(d_i, \phi_i), ~ i = 1, \dots, N,
$
where the variant allele frequency is given by 
$
	\phi_i = (1-e) p_i + e (1-p_i)
$
and $e$ is a sequence read error rate and
$
	p_i = \frac{1}{2} \sum_{k=1}^K \theta_k \bfX_{ki}
$. The parameter $\theta$ is a $K \times 1$ vector denoting the proportion of the observed data sample attributed to each of the $K$ tumor subpopulations whose genotypes are given by a $K \times N$ binary matrix $\bfX$. We specify the prior probabilities over $\theta$ as
$
	\theta_k = \frac{ \gamma_k }{ \sum_{j=1}^K \gamma_j },~k=1,\dots,K,
$ and
$
	\gamma_k  \sim \mathrm{Gamma}(\alpha/K, 1),~k=1,\dots,K,
$
This hierarchical representation is equivalent to a marginal prior distribution $\theta | \alpha \sim \mathrm{Dirichlet}(\alpha/K, \dots, \alpha/K)$ which induces a sparsity forcing values of $\theta$ to be close to zero when $\alpha \leq 1$ allowing us to do automatic model selection for the number of tumor sub-populations. We further specify the prior probabilities over $\bfX$ as
$
	x_{ki} | f_i \sim \mathrm{Bernoulli}( x_{ki}, {f_i} ),~i=1,\dots,N,k=1,\dots,K,
$
and
$
	f_i | f_\alpha, f_\beta \sim \mathrm{Beta}(f_\alpha, f_\beta), ~i=1,\dots,N.
$
Further details for posterior inference for this model and data simulation are given in {\it SI: C Tumor deconvolution with mixture modelling}.

\subsection{Sparse linear regression analysis} Here, we provide a description of the statistical sparse linear regression model 
used in the examples. Suppose a dataset $\{y_i, \bfz_i\}_{i=1}^N$ where $y_i \in \mathbbm{R}$ is the observed 
response and $\bfz_i \in \mathbbm{R}^D$ is the vector of the corresponding covariates.
We can collectively store all responses in a $N \times 1$ vector $\bfy$ (assumed to be normalized to have zero mean) 
and the covariates in a $N \times D$ design matrix $Z$. We further assume that from the total $D$ covariates 
there exist a small unknown subset of relevant covariates 
that generate the response. This is represented by an $D$-dimensional 
unobserved binary vector $\bfX$ that indicates the relevant covariates and follows an 
independent Bernoulli prior distribution, 
$
x_d  \sim \mathrm{Bernoulli}(x_d, \pi_0), \ \ d=1,\ldots,D,
$
where $\pi_0$ is assigned a conjugate Beta prior, $\text{Beta}(\pi_0|\alpha_{\pi_0}, b_{\pi_0})$, and $(\alpha_{\pi_0}, b_{\pi_0})$ 
are hyperparameters. Given $\bfX$, a Gaussian linear regression model takes the form     
$
\bfy = Z_{\bfX} \bfbeta_{\bfX} + \bfeta, \ \ \bfeta \sim \mathcal{N}(\bfzero,\sigma^2 I_N), 
$
where $Z_{\bfX}$ is the $N \times D_{\bfX}$ design matrix, with $D_{\bfX} = \sum_{d=1}^D x_d$, 
having columns corresponding to $x_d=1$ and $\bfbeta_{\bfX}$ is the respective $D_{\bfX} \times 1$ vector 
of regression coefficients. The regression coefficients $\bfbeta_{\bfX}$ and the noise variance 
$\sigma^2$ are assigned a conjugate normal-inverse-gamma prior
$
p(\bfbeta_{\bfX},\sigma^2|\bfX) = \mathcal{N}(\bfbeta_{\bfX}|\bfzero, g (Z_{\bfX}^T Z_{\bfX})^{-1})  \text{InvGa}(\sigma^2|\alpha_{\sigma},b_{\sigma}),
$
where $(g,\alpha_{\sigma}, b_{\sigma})$ are hyperparameters. Notice that the particular choice $g (Z_{\bfX}^T Z_{\bfX})^{-1}$ 
for the covariance matrix , where is $g$ is scalar hyperparameter, corresponds to the so called $g$-prior \cite{Zellner}.
Based on the above form of the prior distributions we can analytically marginalize out the parameters 
$\theta = (\pi_0, \bfbeta_{\bfx}, \sigma^2)$ and obtain the marginalized joint density \cite{bottolo2010}:
$
p(\bfy,\bfX| \cdot ) \propto C  
\left( 2 b_{\sigma} + S(\bfX) \right)^{ - (2 \alpha_{\sigma} + N - 1)/2 }, 
\label{eq:bayesLinReg}
$
where $C = (1 + g)^{ - D_{\bfX}/2 } \Gamma( D_{\bfX} + \alpha_{\pi_0} ) \Gamma( D - D_{\bfX} + b_{\pi_0})$, $S(\bfX) = \bfy^T \bfy  - \frac{g}{1+g} \bfy^T Z_{\bfX} (Z_{\bfX}^T Z_{\bfX})^{-1} 
Z_{\bfX}^T \bfy$ and $\Gamma(\cdot)$ denotes the Gamma function. The hyperparameters of the prior were set to fixed values as follows. The hyperparameters of $\text{InvGa}(\sigma^2|\alpha_{\sigma},b_{\sigma})$ 
were set to $\alpha_{\sigma}=0.1$ and $b_{\sigma}=0.1$ which leads to a vague prior. The scalar hyperparameter for the $g$-prior were chosen to $g=N$ as also used in \cite{bottolo2010}. Finally, the hyperparameters for the  Beta prior, $\text{Beta}(\pi_0|\alpha_{\pi_0}, b_{\pi_0})$, were set to the values $\alpha_{\pi_0}=0.001$ and  $b_{\pi_0}=1$ which favors sparse configurations for the vector $\bfX$. Further details for 
inference for this model and data simulation are given in {\it SI: D Sparse linear regression analysis}.

\subsection{Factorial hidden Markov models} In a typical setting of modelling with FHMMs, the observed sequence $\bfy=(\bfy_1,\ldots,\bfy_N)$ is generated through $K$ binary hidden sequences represented by a $K \times N$ binary matrix $\bfX = (\bfx_1,\ldots,\bfx_N)$. The interpretation of the latter binary matrix is that each row encodes for the presence or absence of a single feature across the observed sequence while each column $x_i$ represents the different features that are active when generating the observation $\bfy_i$. 
Different rows of $\bfX$ correspond to independent Markov chains following 
$
p( x_{ki} | x_{k i-1} ) = (1-\rho_k)^{\mathbb{I}(x_{k i} = x_{k i-1})} \rho_k^{\mathbb{I}(x_{k i} \neq x_{k i-1})},
$
and where the initial state $x_{k 1}$ is drawn from a Bernoulli distribution with parameter $\nu_k$. Each data point $\bfy_i$ is generated conditional 
on $\bfx_i$ through a likelihood model $p(\bfy_i|\bfx_i)$ 
parametrized by $\phi$. For the additive FHMM this likelihood model takes the form 
$
p(\bfy_i\bfx_i) = \mathcal{N}(\bfy_i|\bfw_0 + \sum_{k=1}^K x_{k i} \bfw_k,\sigma^2 I),
$
where $\phi= \{\bfw_0,\ldots, \bfw_K,\sigma^2\}$ are the parameters. 
The whole set of model parameters $\theta = (\phi,\{\rho_k,v_k\}_{k=1}^K)$
determines the joint probability density over $(\bfy,\bfX)$ which is written as 
$
p(\bfy,\bfX|\theta) = \left( \prod_{i=1}^N p(\bfy_i|\bfx_i) \right) \left( \prod_{k=1}^K p(x_{k 1}) \prod_{i=2}^N p( x_{k i} | x_{k i-1} ) \right).
\label{eq:fhmmJoint} 
$
The Hamming ball algorithm follows precisely the iteration in (\ref{eq:u_hbs})-(\ref{eq:X_hbs}). The step (\ref{eq:X_hbs}) when implemented 
by two separate Gibbs step requires sampling from the posterior conditional $p(\bfX|\theta, \bfU, \bfy)$ expressed as
$
p(\bfX|\theta, \bfU, \bfy) \propto \left( \prod_{i=1}^N p(\bfy_i|\bfx_i) \mathbb{I}(\text{d}(\bfx_i,\bfu_i) \leq m) \right) p(\bfX) 
\label{eq:postcondX}
$
where we used that $\mathbb{I}(\bfX \in {\cal H}_m(\bfU)) = \prod_{i=1}^N \mathbb{I}(\text{d}(\bfx_i,\bfu_i)\leq m)$. 
Given that each $\bfx_i$ is restricted to take $M$ values in the neighborhood of $\bfu_i$, exact sampling from the above distribution 
can be done using the FF-BS algorithm in $O(M^2 N)$ time. If we implement step (\ref{eq:X_hbs}) using the M-H joint update, then we need 
to evaluate the $p(\theta, \bfU, \bfy)$ which is simply the normalizing constant of the distribution $p(\bfX|\theta, \bfU, \bfy)$ 
that is obtained by the forward pass of the FF-BS algorithm in $O(M^2 N)$ time. 
Further details for posterior inference for this model and data simulation are given in {\it SI: E Factorial hidden Markov models}.